\documentclass[aps,pra,groupedaddress,floats, reprint]{revtex4-2}

\usepackage[english] {babel}
\usepackage{amsfonts,graphicx,soul,mathrsfs,dsfont}

\usepackage{epstopdf,comment}
\usepackage{xcolor}

\usepackage{amscd, amssymb}

\usepackage{amsmath, nicefrac}
\newcommand{\sign}{\text{sign}}
\usepackage{cancel}

\newcommand{\eu}{\mathrm{e}\mkern1mu}

\usepackage[normalem]{ulem}
\usepackage{accents}

\usepackage{dcolumn}
\usepackage{bm}

\usepackage[colorlinks=true]{hyperref}

\begin{document}

\title{Effective Caldirola-Kanai Model for Accelerating Twisted Dirac States in Nonuniform Axial Fields}

\author{N.V. Filina}%
\email{nvfilina@bk.ru}%
\affiliation{School of Physics and Engineering,
ITMO University, St. Petersburg, Russia 197101}%

\author{S.S. Baturin}%
\email{s.s.baturin@gmail.com}%
\affiliation{School of Physics and Engineering,
ITMO University, St. Petersburg, Russia 197101}%
\date{\today}


\begin{abstract}
We study relativistic twisted (orbital-angular-momentum) states of a massive charged particle propagating through an axially symmetric, longitudinally inhomogeneous solenoid field and a co-directed accelerating or decelerating electric field. Starting from the Dirac equation and using controlled spinless and paraxial approximations, we show that the transverse envelope obeys an effective nonstationary Schr\"odinger equation governed by a Caldirola--Kanai Hamiltonian. The longitudinal energy gain or loss encoded in $f(z)=[E_0-V(z)]^2-m^2$ generates an effective gain or damping rate $\widetilde{\gamma}(z)=\partial_z f(z)/[2f(z)]$ and a $z$-dependent oscillator frequency $\widetilde{\omega}(z)=p_0\Omega(z)/\sqrt{f(z)}$. Exploiting the Ermakov mapping (unitary equivalence of Caldirola--Kanai systems), we obtain a closed-form propagated twisted wave function by transforming the stationary Landau basis. The transverse evolution is controlled by a single scaling function $b(z)$ that satisfies a generalized Ermakov--Pinney equation with coefficients determined by $E_z(z)$ and $B_z(z)$. In the limiting cases of uniform acceleration with $B_z=0$ and of solenoid focusing with negligible acceleration, our solution reduces to previously known analytic results, providing a direct bridge to established models.
\end{abstract}


\maketitle


Twisted (vortex, orbital-angular-momentum, OAM) states are structured wave packets with a well-defined propagation axis and a quantized projection of orbital angular momentum on this axis~\cite{Allen1,Bliokh,UFN}. In cylindrical coordinates, the OAM operator is
$\hat{L}_z=-i\,\partial_\phi$, i.e., the generator of rotations about the $z$ axis. Accordingly, the azimuthal dependence of a twisted state has the form $\exp(i l \phi)$, where $l\in\mathbb{Z}$ is the corresponding eigenvalue of $\hat{L}_z$ (often referred to as the topological charge). Nonzero $l$ is associated with a helical wavefront and a ring-shaped transverse intensity profile.

The acceleration and controlled transport of twisted electrons is a nontrivial problem. Experimentally, twisted electron beams with energies up to $\sim 300~\mathrm{keV}$ and OAM as large as $\sim 1000\hbar$ have been generated in transmission electron microscopes~\cite{Tonomura,Schlattschneider,Agrawal,SCHACHINGER201517,Grillo}. At the same time, twisted states are expected to enable qualitatively new signatures in scattering and radiation processes, including modified angular-momentum selection rules and characteristic changes of radiation patterns in high-energy collisions~\cite{Ivanov2011,IVANOV2022103987,Karlovets+Baturin,PhysRevA.109.012222,Pavlov_neutron,Sheremet}. Extending these studies to higher energies requires dedicated accelerator-compatible schemes for generating, accelerating, and transporting twisted beams.

From the theoretical side, a practical description of twisted-particle evolution in realistic electromagnetic fields remains challenging. Current and near-future experimental programs, including those at JINR~\cite{Dyatlov:25,Dyatlov,dyatlov2025} and proposed efforts in China~\cite{China}, motivate a framework that can treat longitudinally varying fields beyond idealized models. Existing approaches have typically assumed either a uniform magnetic field~\cite{Silenko_NLG,Guzzinati,Schlattschneider}, abrupt field boundaries at solenoid edges~\cite{Karlovets_NLG,Karlovets_short}, or a nonuniform but purely magnetic configuration~\cite{Melkani_Enk,Filina_Baturin_3}. A unified analytic treatment of inhomogeneous electric \emph{and} magnetic fields, while retaining the cylindrical symmetry relevant for twisted states, is still limited.

An important step in this direction was made in Ref.~\cite{Silenko_Electric_field}, where an exact solution was obtained for a relativistic twisted charged particle in a uniform electric field. It was noted that the electric field introduces an additional $z$-dependent ``mass'' factor in the paraxial evolution compared to free space. Such a linear gain or damping can be naturally captured within the Caldirola--Kanai (CK) model~\cite{Caldirola,Kanai}, which we previously employed to solve the propagation problem in a longitudinally inhomogeneous solenoid field~\cite{Filina_Baturin_3}. We emphasize that the effective gain or damping in the transverse CK Hamiltonian arises from the adiabatic longitudinal energy variation and does not imply coupling to an environment.

In our earlier study~\cite{Filina_Baturin_3}, we considered the evolution of an arbitrary quantum state in the magnetic field of an axially symmetric solenoid with circular cross section,
\begin{equation}
\mathbf{B}^T=\left[-\frac{x}{2}\,\partial_z B_z(z)\,;\,-\frac{y}{2}\,\partial_z B_z(z)\,;\,B_z(z)\right],
\end{equation}
and developed an analytic procedure to obtain the propagated wave function for an \emph{arbitrary} longitudinal profile $B_z(z)$. This flexibility enables direct use of experimentally reconstructed field profiles, going beyond simplified piecewise-constant or abrupt-boundary models.

In the present work, we address the relativistic evolution of a single twisted charged particle (electron or proton) in combined inhomogeneous electric and magnetic fields with axial symmetry. We consider solenoid fields as a means of controlling the transverse beam size during propagation, and we include acceleration (or deceleration) by co-directed sections with an arbitrary longitudinal electric-field profile. Starting from the Dirac equation, we apply spinless and paraxial approximations and provide estimates justifying these reductions for typical linac conditions. In the regime of adiabatically varying longitudinal fields, the transverse envelope dynamics reduces to an effective nonstationary Schr\"odinger equation whose Hamiltonian takes the CK form. This equation can be solved by several equivalent analytic methods, including the Ermakov mapping to a stationary oscillator~\cite{QAT2,Filina_Baturin_1} and the construction of Lewis-Riesenfeld invariants~\cite{Melkani_Enk,Lewis1,Lewis2,Lewis}. Using Ermakov framework, we derive a closed-form expression for the propagated twisted wave function in inhomogeneous electric and magnetic fields. As an illustrative example, we model a sequence of solenoids and accelerating sections and show that a twisted electron can recover its initial transverse size and diffraction rate after a finite propagation distance while reaching a higher energy.

Throughout the paper we use natural units, $\hbar=1$ and $c=1$, except when quoting numerical estimates, where we restore $\hbar$ and $c$ for clarity.



A relativistic massive charged particle in an external electromagnetic field is described by the Dirac equation. The accelerating electric field
\begin{align}
    \mathbf{E}^T = \left[0 ; 0 ; E_z(z) \right]
\end{align}
is represented by a static potential $V(z) = q \phi(z)$. Here, $\phi(z)$ is the electric scalar potential, related to the electric field by the standard expression $\mathbf{E} = - \nabla \phi(z)$. For convenience, we set $\phi(0) = 0$. The magnetic field of an axially symmetric solenoid near the axis, according to the Biot--Savart law, is
\begin{align}
    \mathbf{B}^T = \left[- \frac{x}{2} \partial_z B_z(z) ; - \frac{y}{2} \partial_z B_z(z); B_z(z) \right].
\end{align}
To describe this field we introduce the corresponding vector potential $\mathbf{A}(\mathbf{r})$, defined by the relation $\mathbf{B} = \nabla \times \mathbf{A}$. In the Coulomb gauge $\nabla \cdot \mathbf{A} = 0$, the potential takes the following form (for details, see the Supplementary Materials of Ref.~\cite{Filina_Baturin_3}):
\begin{equation}
\label{eq:vecpot}
    \mathbf{A}^{T} =  \left[- \frac{y}{2} B_z(z), \frac{x}{2} B_z(z), 0 \right].
\end{equation}
The Dirac equation in this notation reads
\begin{equation}
\label{eq:dirac}
    \left\{\left(
    \begin{matrix}
    \mathbf{1} & 0 \\
    0 & \mathbf{-1}
    \end{matrix}\right)
    \left[i \frac{\partial}{\partial t} - V(z)\right] - \left(\begin{matrix}
    0 & \boldsymbol{\sigma} \\
    -\boldsymbol{\sigma} & 0
    \end{matrix}\right) \hat{\boldsymbol{\pi}} - m \right\}\left(\begin{matrix} \Phi \\ \chi \end{matrix}\right) = 0,
\end{equation}
where $\hat{\boldsymbol{\pi}} = - i \boldsymbol{\nabla} - q \mathbf{A}$ is the canonical momentum, and $\boldsymbol{\sigma} = (\sigma_x, \sigma_y, \sigma_z)$ is the vector of Pauli matrices. We study stationary solutions to Eq.~\eqref{eq:dirac} with the initial energy $E_0 = \sqrt{p^2_0 + m^2}$, where $p_0$ is the initial momentum of the particle. Thus, we use the standard ansatz for the upper and lower spinors $\Phi$ and $\chi$ in the form~\cite{Chuprikov}
\begin{equation}
    \left[ \begin{matrix} \Phi(\mathbf{r}, t) \\ \chi(\mathbf{r}, t) \end{matrix} \right] = \left[\begin{matrix} \Phi(\mathbf{r}; E_0) \\ \chi(\mathbf{r}; E_0) \end{matrix} \right] \eu^{-i E_0 t}.
\end{equation}
Equation~\eqref{eq:dirac} can be viewed as a system of two coupled equations for the two spinors. We therefore express the lower spinor in terms of the upper spinor,
\begin{equation}
\label{eq:chi}
    \chi = \frac{\boldsymbol{\sigma} \cdot\hat{\boldsymbol{\pi}}}{E_0 - V(z) + m} \Phi.
\end{equation}
After this substitution, and using the Pauli-matrix identity 
\begin{align}
    (\boldsymbol{\sigma} \cdot \mathbf{a})(\boldsymbol{\sigma} \cdot \mathbf{b}) = (\mathbf{a} \cdot \mathbf{b}) + i \boldsymbol{\sigma} \cdot [\mathbf{a} \times \mathbf{b}],
\end{align}
we obtain a closed equation for the upper spinor $\Phi$,
\begin{align}
\label{eq:diracsq}
    \Bigg[ \hat{\boldsymbol{\pi}}^2 - q \boldsymbol{\sigma}\cdot \mathbf{B} - \frac{\partial_z V}{E_0 - V(z) + m} (i \hat{\pi}_z - [\hat{\boldsymbol{\pi}} &\times  \boldsymbol{\sigma}]_z) \Bigg] \Phi \nonumber\\
    &= f(z) \Phi.
\end{align}
Above we introduced the function $f(z)$, which depends on the propagation distance $z$,
\begin{equation}
    f(z) = \left\{[E_0 - V(z)]^2 - m^2\right\}.
\end{equation}

We adopt a spinless approximation. Specifically, in the quadratic Dirac equation~\eqref{eq:diracsq} we neglect the $\boldsymbol{\sigma}$-dependent terms that couple the components of the spinor $\Phi$. We further replace the relevant operators by their characteristic (mean) values, and estimate the contribution involving $\partial_z V$ by comparing its operator norm with the minimal value of $\hat{\boldsymbol{\pi}}^2$ (for accelerating fields, this minimum is attained at $z=0$). Under this estimate, the $\partial_z V$ term is negligible provided that
\begin{align}
    \frac{|\partial_z V| \; \hbar c}{(E_0 + m) \sqrt{E_0^2 - m^2}} \ll 1.
\end{align}
Under the assumptions $E_0 \approx m$ and $|\partial_z V| \approx 10~\mathrm{MeV/m}$, this yields the following condition on the electron kinetic energy,
\begin{align}
\label{eq:ass1}
    E_{kin} \gg \frac{|\partial_z V| \; \hbar c}{2 m} \approx 2 \; \mu \text{eV}.
\end{align}
Thus, for realistic setups, the term proportional to $\partial_z V$ can be neglected.

With these simplifications, Eq.~\eqref{eq:diracsq} reduces to a Helmholtz equation with a variable wave number,
\begin{align}
\label{eq:pi_squared}
    \hat{\boldsymbol{\pi}}^2 \Phi = f(z) \Phi.
\end{align}

To solve Eq.~\eqref{eq:pi_squared} we employ the slowly varying envelope approximation,
\begin{equation}
    \Phi = u(x, y, z) v(z) \eta,
\end{equation}
where $v(z)$ is an oscillatory function of $z$. The spinor $\eta$ is one of the eigenvectors $\eta_+ = \left(\begin{matrix}
    1 \\ 0
\end{matrix}\right)$ or $\eta_- = \left(\begin{matrix}
    0 \\ 1
\end{matrix}\right)$ of the Pauli operator $\hat{\sigma}_z$, with the corresponding eigenvalues $\pm 1$. 

With the vector potential given by Eq.~\eqref{eq:vecpot}, we can separate the longitudinal and transverse dynamics and rewrite Eq.~\eqref{eq:pi_squared} as
\begin{equation}
    \hat{\boldsymbol{\pi}}^2_\perp u v - 2 (\partial_z u) (\partial_z v) - u (\partial^2_{zz}v) = f(z) u v,
\end{equation}
where $\hat{\boldsymbol{\pi}}_\perp$ is the transverse component of the momentum.

Within the slowly varying envelope approximation we neglect $\partial_{zz}u$ compared to $2(\partial_z u)(\partial_z v)/v = 2(\partial_z u)\,\partial_z\!\ln v$.

The resulting system of equations for $u$ and $v$ is
\begin{equation}
\label{eq:v}
    \begin{cases}
        \partial^2_{zz} v + f(z) v = 0, \\
        2 \frac{\partial_z v}{v} \partial_z u = \hat{\boldsymbol{\pi}}^2_\perp u.
    \end{cases}
\end{equation}
We note that the longitudinal part of the wave function $v$ is independent of $u$. At the same time, the logarithmic derivative of $v(z)$ acts as a parameter in the second equation producing and provides effective pumping or dissipation for the transverse part of the wavefunction $u(x,y,z)$.



To solve the first equation in Eq.~\eqref{eq:v} we employ the WKB method. We assume that function $f(z)$ has no turning points. For $f(z) > 0$ (acceleration of the state), the asymptotic solutions are
\begin{equation}
\label{eq:solv}
    v_\pm(z) \approx \frac{1}{\sqrt[4]{f(z)}} \exp \left[ \pm i \int \limits^z_0 \sqrt{f(z')} dz' \right].
\end{equation}
The validity criterion for the WKB approximation is $|\partial_z f| \ll f^{3/2}$, which, in dimensionless form at the point $z = 0$, leads to
\begin{equation}
    \frac{E_0 \hbar c \; |\partial_z V|}{\left(\sqrt{E_0^2 - m^2}\right)^3} \ll 1.
\end{equation}
Under the same assumptions as in estimate~\eqref{eq:ass1} we obtain the lower bound on the kinetic energy,
\begin{align}
\label{eq:ass2}
    E_{kin} \gg (m \hbar c \; |\partial_z V|)^{1/3} \approx 100 \; eV.
\end{align}
This condition is more restrictive than~\eqref{eq:ass1}; nevertheless, we consider electrons with initial kinetic energies well above this threshold.

The two WKB branches in Eq.~\eqref{eq:solv}  correspond to propagation along $\pm z$. Assuming a forward-propagating incident beam and neglecting WKB backreflection (i.e., no branch mixing), we choose the ``$+$'' branch.

For subsequent analysis it is convenient to introduce dimensionless variables (in what follows we omit tildes)
\begin{align}
\label{eq:dim}
    \tilde{z} = \frac{z}{z_0}; \quad \tilde{x} = \frac{x}{\rho_H}; \quad \tilde{y} = \frac{y}{\rho_H},
\end{align}
where we introduce the Rayleigh length $z_0$ and the magnetic length $\rho_H$, and restore dimensional parameters for clarity,
\begin{align}
    z_0 = \frac{p_0 \rho^2_H}{\hbar}; \quad \rho_H = \sqrt{\frac{2 \hbar}{|q| \max{|B_z(z)|}}}.
\end{align}
In this notation, the logarithmic derivative of $v$ is
\begin{equation}
\label{eq:logder}
    \frac{\partial_zv}{v} \approx i p_0 \rho^2_H \sqrt{f(z)}.
\end{equation}
We note that after the rescaling~\eqref{eq:dim} and omission of tildes, $\partial_z$ denotes differentiation with respect to the dimensionless coordinate $\tilde z$.

To solve the second equation in Eq.~\eqref{eq:v}, we substitute the explicit vector potential~\eqref{eq:vecpot}, switch to the dimensionless variables~\eqref{eq:dim}, and obtain
\begin{align}
\label{eq:schrod}
    i \partial_z u = \frac{p_0}{\sqrt{f(z)}} \left[ \hat{H}_\perp - \sign(q) \Omega(z) \hat{L}_z \right] u
\end{align}
with the operators
\begin{align}
\label{eq:Defoper}
    \hat{H}_\perp = \dfrac{\hat{p}^2_\perp}{2} + \Omega^2(z) \dfrac{\hat{x}^2 + \hat{y}^2}{2}, \quad \hat{L}_z = \hat{x} \hat{p}_y - \hat{y} \hat{p}_x.
\end{align}
Here we introduce the dimensionless frequency
\begin{equation}
    \Omega(z) = \frac{B_z(z)}{\max{|B_z(z)|}}
\end{equation}
and use the expression for the logarithmic derivative of $v$ from Eq.~\eqref{eq:logder}. Due to axial symmetry, the following commutator vanishes,
\begin{align}
\forall z',z'' \;\; \left[\frac{p_0}{\sqrt{f(z')}} \hat{H}_{\perp}(z'), \frac{p_0}{\sqrt{f(z'')}} \Omega(z'') \hat{L}_z\right] = 0,
\end{align}
and therefore we can write
\begin{align}
\label{eq:u}
u =\exp\left[\sign(q)\;  i \int\limits_0^z p_0 \frac{\Omega(z')}{\sqrt{f(z')}} dz' \hat{L}_z \right] \widetilde u,
\end{align}
where $\widetilde u$ satisfies a Schr\"{o}dinger-type equation for a two-dimensional harmonic oscillator with damping,
\begin{align}
\label{eq:HO}
    i \partial_z \widetilde u = \frac{p_0}{\sqrt{f(z)}} \hat{H}_{\perp}(z) \widetilde u
\end{align}
with $\hat{H}_{\perp}(z)$ given in Eq.~\eqref{eq:Defoper}.

We note that, unlike the purely magnetic-field setting~\cite{Melkani_Enk,Meng_2025,Filina_Baturin_3}, the presence of the longitudinal electric field rescales the instantaneous Larmor frequency by $\sqrt{f(z)}$. As a result, the rotation dynamics is modified relative to Refs.~\cite{Melkani_Enk,Meng_2025,Filina_Baturin_3} for the same magnetic field profile.

Equation~\eqref{eq:HO} has the same form as a nonstationary Schr\"{o}dinger equation with a Caldirola--Kanai Hamiltonian,
\begin{align}
\label{eq:CKH}
    i \partial_z \widetilde u = \left[\widetilde{w}(z) \dfrac{\hat{p}^2_\perp}{2} +\frac{\widetilde{\omega}^2(z) (\hat{x}^2+\hat{y}^2)}{2 \widetilde{w}(z)}\right] \widetilde u, 
\end{align}
where $\widetilde{w}(z)$ is the Wronskian of the form $\widetilde{w}(z) = \exp\left[-\int \widetilde{\gamma}(z) dz \right]$, and $\widetilde{\gamma}(z)$ is the classical friction coefficient.

The effective parameters of the equivalent nonrelativistic theory are
\begin{align}
    \widetilde{\omega}(z) = \frac{p_0}{\sqrt{f(z)}} &\Omega(z), \quad \widetilde{w}(z) = \frac{p_0}{\sqrt{f(z)}}, \nonumber\\ &\widetilde{\gamma}(z) = \frac{\partial_zf(z)}{2 f(z)}.  
\end{align}
     
In Ref.~\cite{Filina_Baturin_1} we studied twisted solutions to the nonstationary Schr\"{o}dinger equation~\eqref{eq:CKH}. The wave function $\widetilde u$ can be expressed in terms of a single parameter $b(z)$. As discussed in Ref.~\cite{Filina_Baturin_3}, this parameter is directly related to the Twiss $\beta$-function and describes the evolution of the transverse beam size.

We use the unitary equivalence between all systems with Caldirola--Kanai Hamiltonians~\cite{Filina_Baturin_1}. Taking as a reference a harmonic oscillator with $m = 1$, $\omega_0 = 1$, and no dissipation, the parameter $b(z)$ satisfies the Ermakov--Pinney equation~\cite{Ermakov,Pinney},
\begin{align}
\label{eq:Erm}
    &\partial^2_{zz} b +  \frac{\partial_z f(z)}{2 f(z)} \partial_z b + \Omega^2(z) \frac{p^2_0}{f(z)} b = \frac{p^2_0}{f(z)} \frac{1}{b^3}, \nonumber \\
    &b(0) = b_0, \quad \partial_z b(0) = b'_0,
\end{align}
where $b_0$ and $b'_0$ denote the initial transverse beam size and the initial convergence or divergence rate, respectively.

We now switch to transverse polar coordinates $\rho=\sqrt{x^2+y^2}$ and $\phi=\arctan(y/x)$ and use the dimensionless variables introduced above.
To obtain the wave function $\widetilde{u}$, we apply the Ermakov operator~\cite{QAT2,Filina_Baturin_3} to the solutions of the reference system,
\begin{align}
    \label{eq:erm}
    &\widetilde u (\rho,\phi,z) = \hat{\mathcal{E}}\psi_{n, l} (\rho', \phi', z') \\
    &=\frac{1}{b}\psi_{n,l} \left[\frac{\rho}{b},\phi, \int\limits_0^z \frac{p_0}{\sqrt{f(z')}}\frac{dz'}{b^2(z')} \right] \exp\left[{\frac{i \sqrt{f(z)}}{2 p_0}\frac{\partial_z b}{b(z)}\rho^2} \right]. \nonumber
\end{align}
The explicit form of $\psi_{n,l}$ is standard. These basis functions are the stationary Landau states,
\begin{align}
\label{eq:Landau}
    \psi_{n, l} = N_{n, l} \rho^{|l|} \mathcal{L}_{n}^{|l|}\left( \rho^2\right) \exp\left(-\frac{\rho^2}{2} + i l \phi - i \varkappa_{n,l} z\right),
\end{align}
where $\mathcal{L}_{n}^{|l|}$ are generalized Laguerre polynomials, and the corresponding eigenvalues $\varkappa_{n,l}$ are
\begin{align}
    \varkappa_{n,l} = (2n + |l| + 1).
\end{align}
The normalization coefficient $N_{n, l}$ reads
\begin{align}
\label{eq:normc}
    N_{n, l} = \sqrt{\frac{n!}{\pi (n+|l|)!}}.
\end{align}

\begin{widetext}
Combining Eqs.~\eqref{eq:solv}, \eqref{eq:u}, and \eqref{eq:erm}, we obtain the final expression for the upper spinor,
\begin{align}
\label{eq:final}
    \Phi(\rho,\phi,z) = \eta \frac{N_{n,l}}{b} \left(\frac{\rho}{b}\right)^{|l|}\mathcal{L}_{n}^{|l|}\left[\frac{\rho^2}{b^2}\right]\exp\Bigg[-&\frac{\rho^2}{2b^2} + i l \phi\Bigg] \exp\left[{\frac{i \sqrt{f(z)}}{2 p_0}\frac{\partial_z b}{b(z)}\rho^2}-i \varkappa_{n,l} \int\limits_0^z \frac{p_0}{\sqrt{f(z')}}\frac{dz'}{b^2(z')}\right] \nonumber\\
    &\times \frac{1}{\sqrt[4]{f(z)}} \exp\left[ i \int\limits_0^z\sqrt{f(z')}dz' \right] \exp\left[\sign(q) \; i l \int\limits_0^z p_0 \frac{\Omega(z')}{\sqrt{f(z')}} dz' \right] .
\end{align}

\end{widetext}



\begin{figure}[t]
    \centering
     \includegraphics[width=0.48\textwidth]{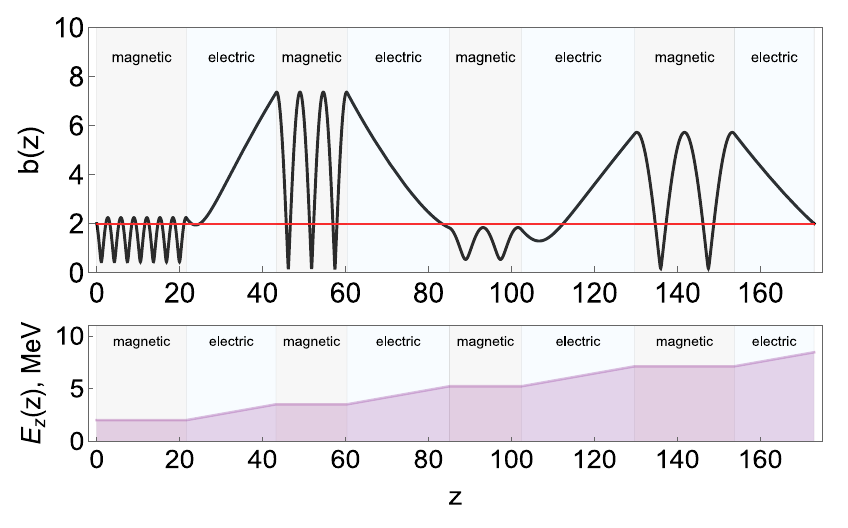}
    \caption{Evolution of the beam width (top panel) and the total relativistic energy (bottom panel) for a twisted electron during acceleration in a system of sequential solenoids and accelerating sections. The red curve corresponds to an initial wave packet size of $b(0)=2$ and $b'(0)=-1$. The energy increases from $2~\mathrm{MeV}$ to $8.45~\mathrm{MeV}$. The maximal magnetic-field amplitude is $0.5~\mathrm{T}$ and the maximal electric-field amplitude is $-10~\mathrm{MV/m}$. The longitudinal coordinate $z$ is measured in units of $z_0 = p_0 \rho^2_H/\hbar \approx 0.025~\mathrm{m}$.}
    \label{Fig:2}
\end{figure}

Particle accelerators typically consist of multiple sections with distinct electromagnetic-field configurations, making models based on constant homogeneous fields impractical. Nevertheless, a number of exact solutions exist in the literature for constant homogeneous magnetic fields, uniform electric fields, and other idealized configurations~\cite{Silenko_Electric_field,Silenko_NLG,Melkani_Enk,Karlovets_NLG}. The Ermakov-Pinney equation provides a more flexible framework, allowing one to go beyond such simplifications and incorporate realistic experimental field profiles. As we show in the Supplementary Materials~\cite{SuppMat}, its general solution can be constructed from two solutions to the associated Hill equation. However, the Hill equation itself does not admit a general closed-form solution for arbitrary coefficients. For this reason, it is natural at this stage to substitute specific field configurations directly into Eq.~\eqref{eq:Erm}. The resulting formalism yields a universal description of twisted solutions to the Dirac equation for arbitrary arrangements of solenoids and longitudinal electric fields.

As an illustrative example, we consider a configuration with coaxial sequential solenoids and acceleration sections. In the top panel of Fig.~\ref{Fig:2} we solve the Ermakov-Pinney equation~\eqref{eq:Erm} numerically for piecewise-constant magnetic and electric fields, obtaining the evolution of the beam width in units of the magnetic (Landau) length. We choose the set of sections such that after approximately $4.4~\mathrm{m}$ the twisted state recovers its initial beam width and diffraction rate. The bottom panel shows the particle's total energy, which increases from $2~\mathrm{MeV}$ to $8.45~\mathrm{MeV}$ during propagation. This example illustrates the feasibility of controlling diffraction during the acceleration of a twisted massive charged particle.



In this work, we developed a universal theoretical framework for the evolution of a relativistic twisted charged particle in spatially inhomogeneous, longitudinal electric fields and axially symmetric solenoid magnetic fields. Starting from the Dirac equation, we provided a transparent reduction to an effective Schr\"odinger-type system for the transverse envelope and supplied estimates showing that the underlying spinless and paraxial approximations are well satisfied for typical linac conditions.

A key outcome is that the reduced transverse dynamics is governed by an effective Caldirola--Kanai Hamiltonian with a $z$-dependent gain or damping term and a renormalized oscillator frequency. This structure follows from the adiabatic regime relevant for accelerator-type fields, in which the longitudinal energy varies slowly,
$|\partial_z f(z)| \ll f^{3/2}(z)$, so that the longitudinal energy gain or loss encoded in $f(z)$ enters the transverse dynamics as an effective ``dissipation'' or ``pumping'' term. Importantly, this term does not imply coupling to an environment; it is a consequence of the controlled reduction of the unitary relativistic dynamics to an envelope equation.

Motivated by the hint of Ref.~\cite{Silenko_Electric_field}, we generalized the Ermakov approach to treat simultaneously $z$-dependent electric and magnetic fields, thereby extending our previous analytic framework for cylindrically symmetric solenoid profiles~\cite{Filina_Baturin_3} to include longitudinal acceleration or deceleration. Our reduction to an effective nonrelativistic transverse dynamics is obtained directly from the Dirac equation within controlled spinless and paraxial approximations, without invoking a Foldy--Wouthuysen transformation, providing an independent route to the same class of envelope equations. We constructed the propagated twisted wave function by applying an Ermakov transformation to the stationary Landau basis, yielding the closed-form expression for the upper spinor component in Eq.~\eqref{eq:final}. The entire transverse evolution is controlled by a single scaling function $b(z)$ satisfying the generalized Ermakov--Pinney equation~\eqref{eq:Erm}, whose coefficients are determined directly by the experimental field profiles $E_z(z)$ and $B_z(z)$.

We illustrated the formalism for a sequence of coaxial solenoids and accelerating sections. We demonstrated, by direct numerical solution of Eq.~\eqref{eq:Erm}, that acceleration with controlled diffraction is achievable by an appropriate choice of element lengths, allowing the twisted state to recover its initial transverse width and diffraction rate after a finite propagation distance while increasing its energy.

We also emphasize that the present construction is not restricted to monotonic longitudinal potentials: it remains applicable for general, nonmonotonic profiles that admit turning points. In that case, the longitudinal part of the wave function can be treated within the standard quasiclassical (WKB) framework, including the usual connection formulas across turning points, which enables the determination of longitudinal bound states when confinement is present and, in turn, determines the transverse mode structure for the corresponding trapped (quantized) longitudinal motion.

Overall, the present results provide a practical route to incorporate longitudinal acceleration into the previously developed analytic model of quantum-state evolution in cylindrically symmetric magnetic fields~\cite{Filina_Baturin_3}, and offer a convenient tool for designing and interpreting accelerator-compatible transport of twisted charged-particle beams.


\begin{acknowledgments}

The work was supported by the Priority 2030 Academic Program.

\end{acknowledgments}




\clearpage
\begin{widetext}
\section*{Supplemental Material}

\section{Analytical solution to the Ermakov equation in an arbitrary electric field \label{app:1}}
As we have noticed in the main text, the upper spinor is parametrized by the $b(z)$ function that shows the transverse size evolution. Next, we analyze the Ermakov-Pinney equation (Eq.(30) of the main text). We start from the limit with zero magnetic field and an arbitrary $z$-dependent electric field. In this case the Ermakov equation can be written in the form
\begin{equation}
    2 f(z) \partial^2_{zz} b + \partial_z f(z) \partial_z b = \frac{2 p^2_0}{b^3}.
\end{equation}
We can integrate this differential equation once and get
\begin{equation}
    f(z) (\partial_z b)^2 - f(0) (b'_0)^2 = 2 p^2_0 \int\limits^b_{b_0} \frac{d \tilde{b}}{\tilde{b}^3} = - \frac{p^2_0}{b^2(z)} + \frac{p^2_0}{b^2_0}. 
\end{equation}
By the definition $f(0) = p^2_0$, so we denote the constant term as $c = \frac{1}{b^2_0} + (b'_0)^2$ and express the derivative of the function $b$
\begin{equation}
    \partial_z b = \sign(b'_0) \; p_0 \sqrt{\frac{1}{f(z)} \left( c - \frac{1}{b^2(z)} \right) }.
\end{equation}
The next step is to separate variables
\begin{equation}
    \int\limits^b_{b_0} \frac{\tilde{b} d \tilde{b}}{\sqrt{c \tilde{b}^2(z) - 1}} = \sign(b'_0) \; p_0 \int\limits^z_0 \frac{d \tilde{z}}{\sqrt{f(\tilde{z})}}. 
\end{equation}
After evaluation of the left side of the equality, we obtain
\begin{equation}
    \int\limits^{c b^2 - 1}_{c b^2_0 - 1} \frac{dx}{2 \sqrt{x}} = \sqrt{c b^2(z) - 1} - \sqrt{c b^2_0 - 1} = \sign(b'_0) \; c p_0 \int\limits^z_0 \frac{d \tilde{z}}{\sqrt{f(\tilde{z})}}.
\end{equation}
For an arbitrary electric field, we can find the evolution of the transverse size of the state in integral representation
\begin{equation}
\label{eq:belf}
    b(z) = \frac{1}{\sqrt{c}} \sqrt{1+ \left(\sqrt{c b^2_0 - 1} + \sign(b'_0) \; c p_0 \int\limits^z_0 \frac{d \tilde{z}}{\sqrt{f(\tilde{z})}}\right)^2}.
\end{equation}
We now consider the special case of a homogeneous electric field that was discussed in paper \cite{Silenko_Electric_field}. The function $f(z)$ of the dimensionless coordinate $z$ has the form
\begin{equation}
    f(z) = \left[(E_0 + e E_z p_0 \rho^2_H z)^2 - m^2\right] = p^2_0 \left( 1 + 2 \tilde{K}_1 z + \tilde{K}^2_2 z^2 \right),
\end{equation}
where we introduce dimensionless coefficients for shorter notation
\begin{align}
    \tilde{K}_1 = \frac{E_0 e E_z \rho^2_H}{p_0}; \quad \tilde{K}_2 = e E_z \rho^2_H.
\end{align}
After the integral evaluation in Eq. \eqref{eq:belf} we obtain the evolution of the transverse size of the wave packet in homogeneous electric field
\begin{align}
    b(z) = \frac{1}{\sqrt{c}} \sqrt{1+ \left\{\sqrt{c b^2_0 - 1} + \sign(b'_0) \; \frac{2 c}{\tilde{K}_2} \text{arctanh}\left[ \frac{\tilde{K}_2 \left(\sqrt{1 + 2 \tilde{K}_1 z + \tilde{K}^2_2 z^2} - 1 - \tilde{K}_1 z\right)}{\tilde{K}^2_2  z - \tilde{K}_1 \left(\sqrt{1 + 2 \tilde{K}_1 z + \tilde{K}^2_2 z^2} - 1\right)} \right]\right\}^2}.
\end{align}
We use the connection between hyperbolic tangent and natural logarithm in form $\text{arctanh} \; x = \dfrac{1}{2} \ln{\dfrac{1+x}{1-x}}, \; \; |x| < 1$ and obtain in previous notation 
\begin{align}
\label{eq:bmain}
    b(z) = \frac{1}{\sqrt{c}} \sqrt{1+ \left\{\sqrt{c b^2_0 - 1} + \sign(b'_0) \; \frac{c p_0 z}{V(z)} \ln\left[ \frac{E_0 + p_0}{E_0 - p_0} \frac{\sqrt{f(z)} - p_0 + V(z)}{\sqrt{f(z)} - p_0 - V(z)} \right]\right\}^2}.
\end{align}

The same logic could be applied to the evolution in the homogeneous magnetic field and the inhomogeneous electric field. The corresponding differential equation can be analytically solved through the presented algorithm.

\end{widetext}

\section{Homogeneous electric field cross-check}
In Ref.~\cite{Silenko_Electric_field} the authors consider the case of homogeneous electric field and derive the expression for the beam width $w(z)$ [Eq.~(18) therein]:
\begin{align}
\label{eq:wsil}
    w(z) &= w_0 \sqrt{\left[ 1 + \frac{2 A(z) w'_0}{K_2 w_0}^2 + \frac{16 A^2(z)}{k^2_0 K^2_2 w^4_0} \right]},  \\
    A(z) &= \text{arctanh}\; \left\{ \frac{K_2}{2 K_1 + K^2_2 z} \left[ \frac{k(z)}{k_0} - 1 \right] \right\}. \nonumber
\end{align}

To compare approaches, we express their parameters in our notation (we return tildes above the dimensionless coordinates for clarity)
\begin{align}
    k(\tilde{z} z_0) = k_0 \sqrt{1 + 2 K_1 \tilde{z} z_0 + K^2_2 \tilde{z}^2 z^2_0} = \sqrt{f(\tilde{z})},
\end{align}
where
\begin{align}
    k_0 = p_0; \quad K_1 = \frac{E_0 e E_z}{p^2_0}; \quad K_2 = \frac{e E_z}{p_0}.
\end{align}
In addition, due to the different notation in the definition of Landau states the $w$ and $b$ are the same up to the multiplier
\begin{align}
    w(\tilde{z} z_0) = \sqrt{2} \rho_H b(\tilde{z}).
\end{align}
In particular, the initial parameters of the beam are connected as (remember that primes near $w$ and $b$ mark derivatives over $z$ and $\tilde{z}$ respectively)
\begin{align}
    w_0 = \sqrt{2} \rho_H b_0; \quad w'_0 = \frac{\sqrt{2}}{p_0 \rho_H} b'_0.
\end{align}

\begin{figure}[t]
    \centering
    \includegraphics[width=0.48\textwidth]{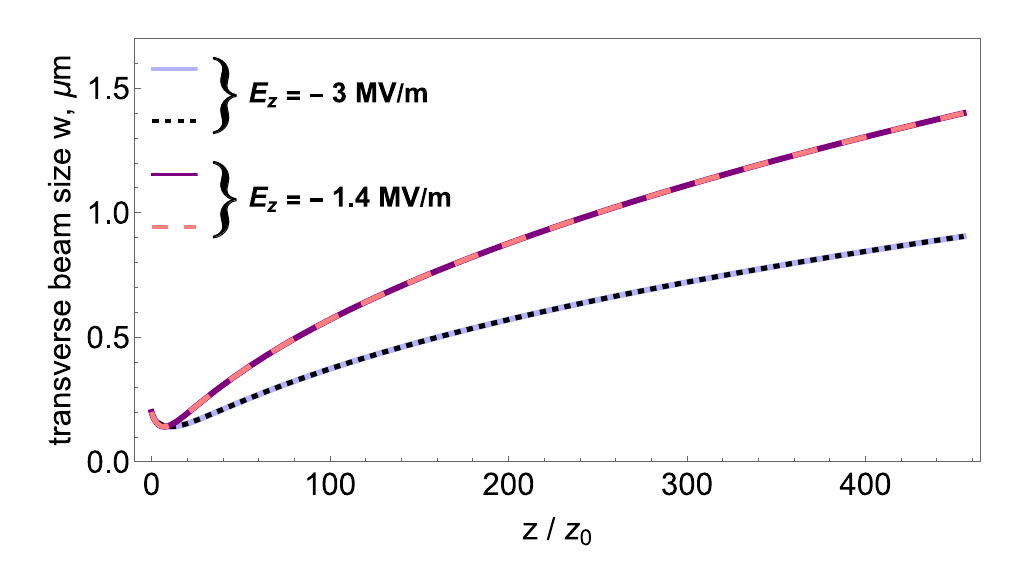}
    \caption{Evolution of the twisted electron beam width in a homogeneous electric field with zero magnetic field. The parameters are $w_0 = 0.2 \; \mu m$, $\partial_{\tilde{z}} w_0 = - 2/(p_0 w_0)$, $p_0 = 0.02 \; MeV$, $q = - |e|$. Blue and dashed black curves correspond to the electric field $E_z = -1.4 \; MV/m$, purple and dashed rose curves correspond to $E_z = -3 \; MV/m$. The longitudinal coordinate $z$ is measured in units of $z_0 = p_0 \rho^2_H/\hbar \approx 0.025$ m.}
    \label{Fig:1}
\end{figure}


Figure \ref{Fig:1} illustrates the equivalence of two approaches for describing the evolution of the beam parameters. The solid curves are obtained from Eq.~\eqref{eq:bmain} multiplied by a factor of $\sqrt{2} \rho_H$, while the dashed curves correspond to Eq.~\eqref{eq:wsil} as a function of the dimensionless longitudinal coordinate $\tilde{z}$. For this specific field configuration, the scaling parameter is chosen to be the same as for a constant magnetic field with a magnitude of $0.5$ T. The two solutions are also compared via their pointwise norm, which differs only at the level of machine precision. This agreement provides a nontrivial cross-check in the homogeneous-electric-field limit; for general profiles, the dynamics is captured by Eq.~\eqref{eq:belf} and, in the presence of a magnetic field, by the full Ermakov--Pinney equation in the main text.

\section{Connection between solutions to the Hill equation and solutions to the Ermakov-Pinney equation \label{app:2}}

In the main text, we obtain the Ermakov-Pinney equation in the form
\begin{multline}
\label{eq:Ermapp}
    b'' +  \widetilde{\gamma}(z) b' + \widetilde{\omega}^2(z) b = \frac{\widetilde{w}^2(z)}{b^3}, \\
    b(0) = b_0, \quad b'(0) = b'_0, \quad \widetilde{w}(0) = 1
\end{multline}
where primes denote derivatives with respect to $z$.

In this section, we show that the solution $b(z)$ can be constructed from two linearly independent solutions $s(z)$ and $t(z)$ to the Hill equation
\begin{multline}
    s'' + \widetilde{\gamma}(z) s' + \widetilde{\omega}^2(z) s = 0, \\
    s(0) = 0, \quad s'(0) = 1/b_0;
\end{multline}
\begin{multline}
    t'' + \widetilde{\gamma}(z) t' + \widetilde{\omega}^2(z) t = 0, \\
    t(0) = b_0, \quad t'(0) = b'_0
\end{multline}

with the Wronskian $\widetilde{w}(z) = s't - st'$.

Indeed, the explicit transformation is 
\begin{equation}
    b = \sqrt{s^2 + t^2}.
\end{equation}

We verify this by direct calculation:
\begin{align*}
    &b' = \frac{ss' + tt'}{\sqrt{s^2 + t^2}}, \\
    &b'' = \frac{ss'' + s'^2 + tt'' + t'^2}{\sqrt{s^2 + t^2}} - \frac{(ss'+tt')^2}{\left(\sqrt{s^2 + t^2}\right)^3} \nonumber\\
    &= \frac{s''s(s^2+t^2) + t''t (t^2 + s^2) + s'^2 t^2 + s^2 t'^2 -  2 s s' t t'}{\left(\sqrt{s^2 + t^2}\right)^3}.
\end{align*}
\\
After substitution in the Ermakov-Pinney equation \eqref{eq:Ermapp} we obtain
\begin{align*}
    &\Big[-(\widetilde{\gamma}s'+\widetilde{\omega}^2 s) s (s^2+t^2) - (\widetilde{\gamma}t'+\widetilde{\omega}^2 t) t (s^2+t^2) \\
    &+ s'^2 t^2 + s^2 t'^2 -  2 s s't t' + \widetilde{\gamma} (ss'+tt')(s^2+t^2) \\
    &+ \widetilde{\omega}^2 (s^2+t^2)^2 \Big] \frac{1}{\left(\sqrt{s^2 + t^2}\right)^3} = \frac{\widetilde{w}^2}{\left(\sqrt{s^2 + t^2}\right)^3}.
\end{align*}

All terms proportional to $\widetilde{\gamma}$ and $\widetilde{\omega}$ cancel. Finally, we obtain
\begin{align}
    s'^2 t^2 + s^2 t'^2 - 2 s s' t t' = (s't - s t')^2 = \widetilde{w}^2,
\end{align}
which is precisely the definition of the Wronskian squared.

The uniqueness theorem for ordinary differential equations establishes that the solution to the Ermakov-Pinney equation with initial conditions can be reframed as the solution to a pair of Hill equations with specified initial conditions.




\bibliographystyle{apsrev4-2}
\bibliography{refs}


\end{document}